\documentclass[aip,rsi,numerical,reprint]{revtex4-1}
\usepackage{graphicx} 
\usepackage{dcolumn}
\usepackage{color}
\usepackage{amsmath}
\usepackage{amsfonts}
\usepackage{amssymb}

\usepackage{bm}
\usepackage{txfonts}
\newcommand{\sub}[1]{\ensuremath{_{\mathrm {#1}}}}

\newcommand{\degc}{$^{\circ}$C}
\renewcommand{\deg}{^{\circ}}
\newcommand{\Tc}{T\sub{c}}

\newcommand{\chiac}{\chi\sub{ac}}
\newcommand{\Ohm}{\ensuremath{\Omega}}
\newcommand{\sro}{Sr\sub{2}RuO\sub{4}}
\begin{document}

\title{Compact AC Susceptometer for Fast Sample Characterization down to 0.1~K}

\author{Shingo~Yonezawa}
\affiliation{Department of Physics, Graduate School of Science, Kyoto University, Kyoto 606-8502, Japan}
\author{Takumi~Higuchi}
\affiliation{Department of Physics, Graduate School of Science, Kyoto University, Kyoto 606-8502, Japan}
\author{Yusuke~Sugimoto}
\affiliation{Department of Physics, Graduate School of Science, Kyoto University, Kyoto 606-8502, Japan}
\author{Chanchal~Sow}
\affiliation{Department of Physics, Graduate School of Science, Kyoto University, Kyoto 606-8502, Japan}
\author{Yoshiteru~Maeno}
\affiliation{Department of Physics, Graduate School of Science, Kyoto University, Kyoto 606-8502, Japan}

\email{yonezawa@scphys.kyoto-u.ac.jp}

\date{\today}

\begin{abstract}
We report a new design of an AC magnetic susceptometer compatible with the Physical Properties Measurement System (PPMS) by Quantum Design, as well as with its adiabatic demagnetization refrigerator option. 
With the elaborate compact design, the susceptometer allows simple and quick sample mounting process. 
The high performance of the susceptometer down to 0.1~K is demonstrated using several superconducting and magnetic materials.
This susceptometer provides a method to quickly investigate qualities of a large number of samples in the wide temperature range between 0.1 and 300~K.
\end{abstract}

\maketitle



\section{Introduction}\label{sec:intro}

In material science and particularly in solid state physics, finding materials exhibiting novel phenomena is indispensable for evolution of research fields.~\cite{Bednorz1986,Maeno1994,Tokura1999.JMagMagMat.200.1}
On the other hand, improvement of sample quality of a known material is a key to explore unresolved issues.~\cite{Mackenzie2003RMP,Borzi2007.Science.315.214,Doiron-Leyraud2007.Nature.447.565,Aoki2014.JPhysSocJpn.83.061101}
For example, in case of unconventional superconductors, sample quality can be evaluated by measuring their critical temperature $\Tc$ and the superconducting (SC) volume fraction.
For both types of researches, one needs to evaluate properties of every obtained sample.
Thus, methods to quickly check properties of obtained samples largely accelerate research progress.

AC susceptibility $\chiac$ is an excellent technique for both search and sample-quality-check of SC and magnetic materials.
Firstly, it takes very short time for measurement preparation because it is a contact-less technique.
Secondly, in addition to transition temperatures, $\chiac$ also provides other rich information, such as the volume fraction of the SC or magnetic ordering, changes in the magnetic response inside the ordered phases, etc.

Here, we report a newly designed susceptometer which fits the widely-used commercial refrigerator Physical Properties Measurement System (PPMS) manufactured by Quantum Design.
The AC susceptibility of a sample is measured with the ordinary mutual inductance method using a lock-in amplifier.
Our new susceptometer is compatible not only with the ordinary sample chamber of the PPMS but also with the recently released adiabatic demagnetization refrigerator (ADR) option.~\cite{Aoki2014.JPhysSocJpn.83.061101,ADR.PressRelease}.
Using paramagnetic salt (chromium alum), the ADR option enables us to cool down the sample from room temperature to below 0.1~K within two hours.
Thus, in particular, the combination between the ADR option and AC susceptibility would provide a powerful tool to investigate the sample quality quickly down to 0.1~K without being annoyed by complicated preparation processes such as attaching electrical leads to the sample.

\section{Design and usage of the Susceptometer}

\begin{figure}[tb]
\begin{center}
\includegraphics[bb=24.094200 21.852500 868.911000 1177.190000,width=8.5cm]{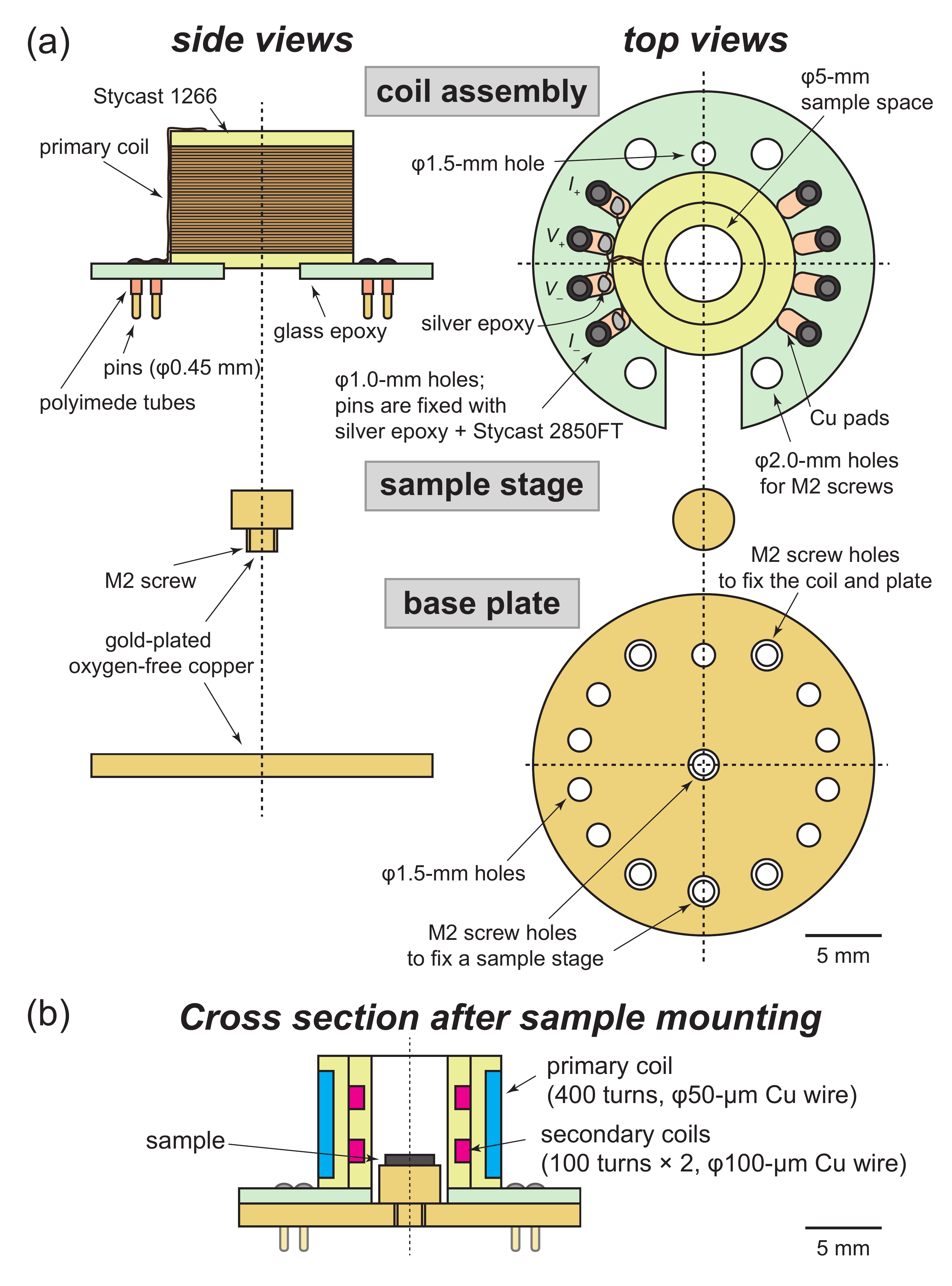}
\caption{Schematic description of the susceptometer. 
(a) Description of each part. The coil assembly consists of a susceptometer coil wound onto epoxy bobbin placed on a glass-epoxy plate. 
The electric pins are also fixed to this assembly.
The base plate made of gold-plated oxygen-free copper has several screw holes as well as through holes which the pins go through.
The sample stage, also made of gold-plated oxygen-free copper, can be fixed to the base plate.
(b) Cross-sectional view of the susceptometer when a sample is mounted.
The blue and red regions indicate the primary and secondary coils, respectively.
\label{fig:coil}
}
\end{center}
\end{figure}

As presented in Fig.~\ref{fig:coil}(a), the susceptometer consists of two major parts, the coil assembly and the base plate.
The coil assembly consists of the susceptometer coil mounted on a thin insulating plate with electrical pins.
The susceptometer coil consists of the primary coil on the outside and the secondary coils inside, as described in the cross-sectional view in Fig.~\ref{fig:coil}(b).
The primary coil was made of 400 turns of insulated copper wire ($\phi50~\muup$m).
The AC field generated by the primary coil is approximately 0.033~T/A at the sample position (see Sec.~\ref{sec:field-distribution}).
The secondary coils are a series of counter-wound coils with 100 turn each made of insulated copper wire ($\phi100~\muup$m).
For both coils, the coil bobbins were machined from rods of epoxy (Stycast 1266, Henkel Ablestik Japan Ltd.).
The secondary coil bobbin was fixed inside the primary coil with varnish (GE 7301, General Electric).
For the plate below the coil, we used a thin glass epoxy plate with copper foil on the top surface.
This plate was machined according to Fig.~\ref{fig:coil}(a) and the copper foil was etched out with FeCl\sub{3} except for small pads around the eight holes for the electric pins.
The pins were taken from an electric socket (ME-10-10, MAC8) and were fixed to the plate with silver epoxy (H20E, Epoxy Technology Inc.) to achieve electrical contacts.
The silver epoxy was cured in air at 100\degc.
Subsequently, we put insulating epoxy (Stycast 2850FT, Henkel Ablestik Japan Ltd.) to improve mechanical strength of the pins.
We fixed short polyimide tubes at the root of the pins to avoid short circuit to the ground.
Finally, the coil was glued to the glass-epoxy plate with varnish, and the wire of the coils were connected to the copper pads with the silver epoxy, which was cured again at 100\degc.

The base plate (Fig.~\ref{fig:coil}(a)) is made of oxygen-free copper.
Eight through holes were made so that the electric pins of the coil assembly can go through the plate.
We made four M2 screw holes for connection with the coil assembly.
Two additional screw holes were prepared to fix a variety of sample stages to the base plate.
After machined, it was gold plated to achieve good thermal connection even at sub-Kelvin temperature ranges and to avoid surface oxidization.

\begin{figure}[tb]
\begin{center}
\includegraphics[bb=8.503890 9.007320 482.940000 342.017000,width=8.5cm]{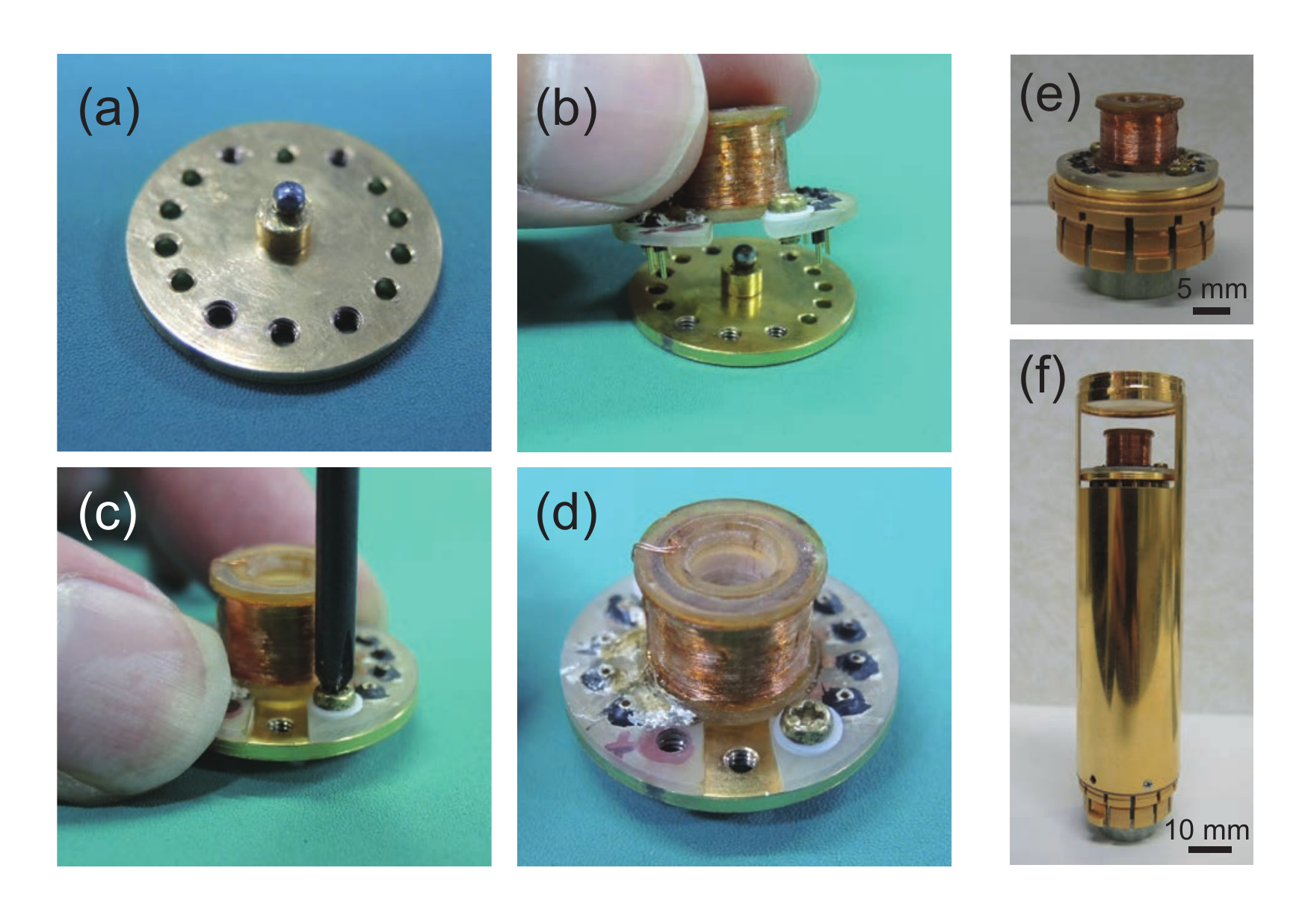}
\caption{Photos describing the preparation process of our susceptometer:
(a) Mount the sample onto the sample stage. 
(b) Put the coil assembly over the plate.
(c) Tighten the screws to fix the coil assembly to the base plate.
(d) Completed assembly. 
The processes (a)-(d) take only a few minutes.
The panels (e) and (f) are photos of the susceptometer placed on an adapter for ordinary PPMS sample chamber and on the ADR option, respectively.
\label{fig:mounting}
}
\end{center}
\end{figure}

We made a small sample stage ($\phi 4$~mm) with gold-plated oxygen-free copper.
This stage is fixed inside the coil by using the screw hole at the center of the base plate.
One can also use a sample stage with `L' shape using thin copper plate, e.g., when the sample should be fixed vertically.
In this case, one can use the other screw hole located near the edge of the plate (see Fig.~\ref{fig:coil}(a)) to fix the stage.
Note that the opening of the glass-epoxy plate allows an easy access to this screw hole.

Figure~\ref{fig:mounting} describes the sample mounting process.
One first put the sample on top of the sample stage with a small amount of grease (N Grease, Apiezon) as shown in Fig.~\ref{fig:mounting}(a). 
After that, one fix the coil assembly to the base plate with M2 screws (Figs.~\ref{fig:mounting}(b) and (c)), to finish the mounting process (Fig.~\ref{fig:mounting}(d)).
Owing to the design enabling one to completely separate the coil and the base plate, one can easily access the sample space.
Typically, it only takes a few minutes for this mounting process.
Since there is no floating wire in the susceptometer, the risk of damaging susceptometer wire is very small even for untrained users.

\section{Electronics}

\begin{figure}[tb]
\begin{center}
\includegraphics[bb=0.000000 0.000000 563.757000 444.508000,width=8.5cm]{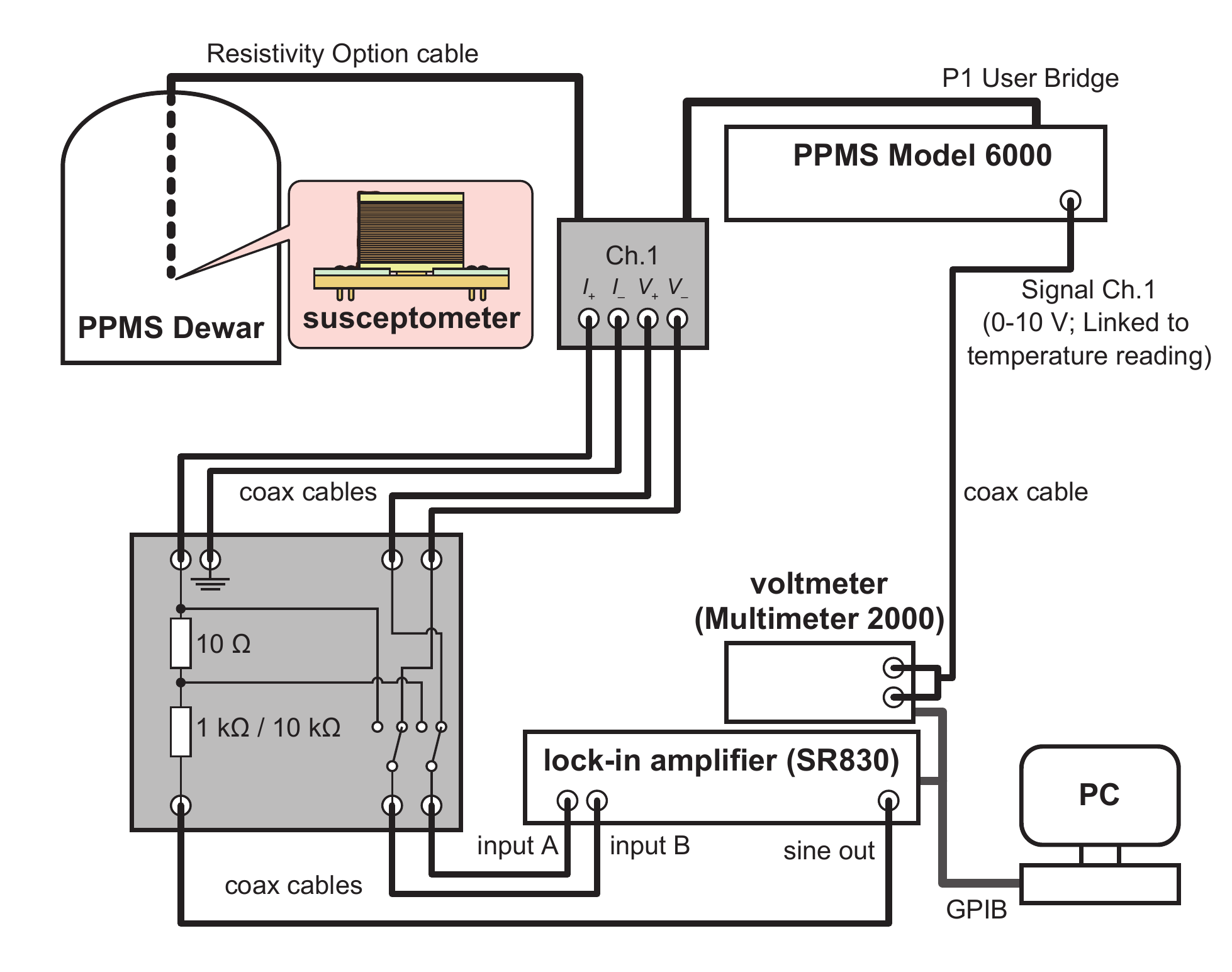}
\caption{Schematic description of the electronics used in this study. In addition to the susceptometer, the boxes shown in gray color are designed and fabricated by the authors.
\label{fig:electronics}
}
\end{center}
\end{figure}

External electronics to measure the temperature dependence of the susceptometer response is connected as described in Fig.~\ref{fig:electronics}.
We insert a box between the Resistivity Option cable and the Physical Properties Measurement System (PPMS) Model 6000 to have connections to the susceptometer placed inside the dewar of the PPMS.
The four wires of Channel 1 of the PPMS are used for the susceptometer: $I_+$ and $I_-$ terminals are used to supply current to the primary coil, while $V_+$ and $V_-$ terminals are to measure the voltage produced by the secondary coil.
Note that Channel 3 is used for thermometry in the adiabatic demagnetization refrigerator (ADR) and ${}^3$He options. 
Thus, wires for Channel 3 should be connected back to the ``P1 User Bridge'' port of Model 6000.

The susceptometer response is detected with the lock-in amplifier (SR830, Stanford Research Systems), with which one can measure the in-phase ($V_x$) and out-of-phase ($V_y$) components of the AC voltage of selected frequency.
For measurements below 2~K, we used ``auto offset'' and ``expansion'' functions of SR830 in order to maximize the voltage resolution. Use of these functions are not recommended for higher temperature measurements, where the temperature-dependent background signal may cause overload of the amplifier.
The excitation current to the primary coil is supplied from the ``sine out'' terminal of SR830, through a resistor of 1~k\Ohm\ or 10~k\Ohm, fixed in another box.
The resisters are necessary to achieve electric current (nearly) independent of temperature and magnetic-field conditions.
The precise values of the phase and amplitude of the current are determined by monitoring the voltage across the standard register of $10~\Ohm$ before sample measurements.
Voltage signal of the range $-10$--10~V proportional to the temperature reading of PPMS is supplied from the ``Signal Ch.1'' terminal of Model 6000 and read by a voltmeter (Multimeter 2000, Keithley).

\section{Performance of the Susceptometer}

By using the susceptometer with a PPMS, we measured the susceptometer responses of several superconducting and magnetic materials.
In the following, we present background-subtracted data $\varDelta V_x$ and $\varDelta V_y$. 
Details of the background subtraction are described in Appendix.

\begin{figure}[tb]
\begin{center}
\includegraphics[bb=0.000000 0.000000 360.000000 252.000000,width=8.5cm]{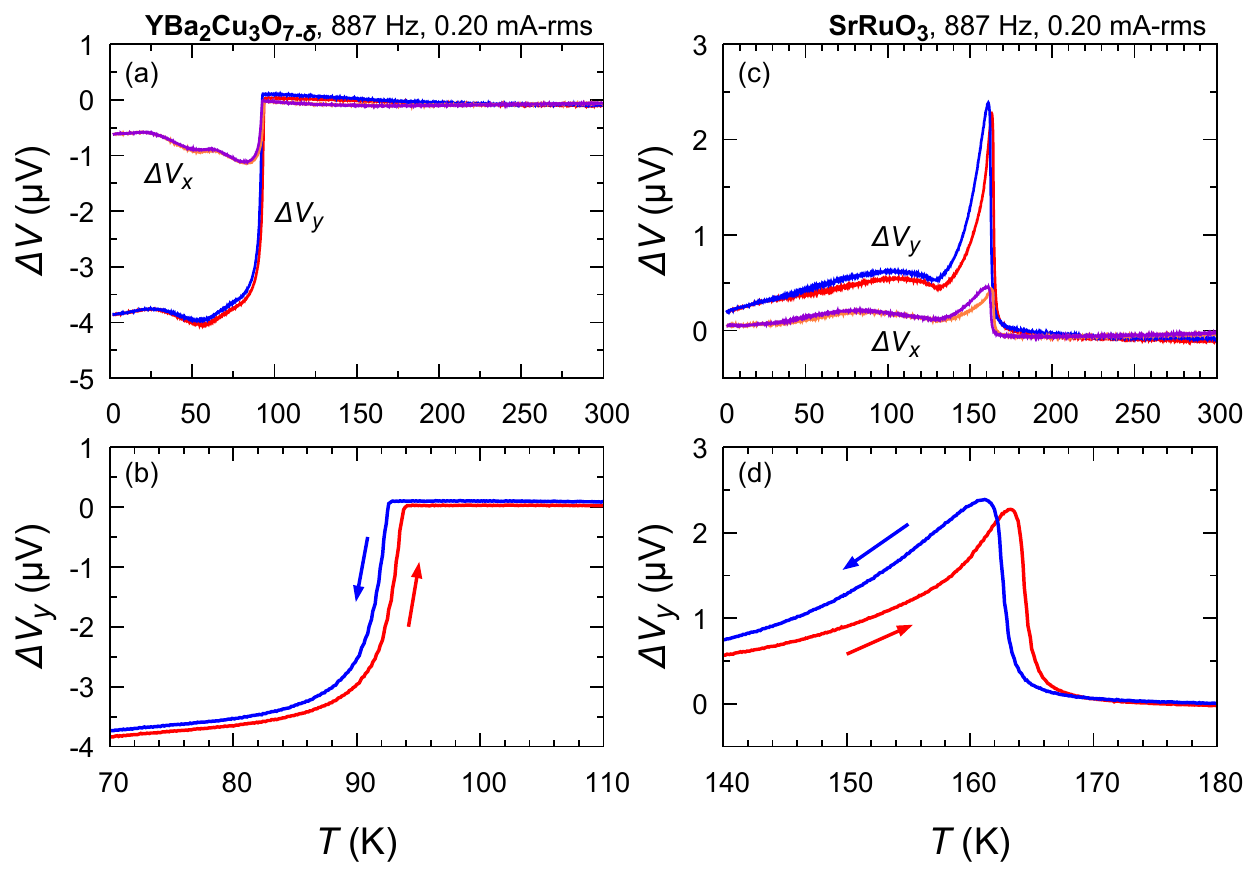}
\caption{Susceptometer responses of a polycrystalline YBCO sample (43.50~mg) and a polycrystalline SrRuO\sub{3} sample (43.38~mg) measured with the AC current of 0.20~mA-rms ($\sim 6.6$~$\muup$T-rms) and 887~Hz.
Background-subtracted $\varDelta V_y$ and $\varDelta V_x$ data of YBCO and SrRuO\sub{3} are presented in (a) and (c), respectively.
Enlarged view of $\varDelta V_y$ near transition temperatures are in (b) and (d). The arrows in (b) and (d) indicate the direction of temperature sweeps.
\label{fig:YBCO_SRO113}
}
\end{center}
\end{figure}

As the first set of examples, we present the temperature dependence of the susceptometer response of polycrystalline samples of the high-$\Tc$ superconductor YBa\sub{2}Cu\sub{3}O$_{7-\delta}$ (YBCO) with $\delta\sim 0.1$ and the itinerant ferromagnet SrRuO\sub{3} in Fig.~\ref{fig:YBCO_SRO113}, measured with the ordinary ${}^4$He cryostat of PPMS (see Fig.~\ref{fig:mounting}(e)) and at $H\sub{DC} = 0$.
The samples were prepared with ordinary solid-state reactions.
For YBCO, a clear diamagnetic signal originating from superconductivity was observed below $\Tc = 93.2\pm0.6$~K, which agrees with the reported value of $\Tc\sim 93$~K.~\cite{Wu1987.PhysRevLett.58.908} 
In the SC state, we observed additional features (Fig.~\ref{fig:YBCO_SRO113}(a)), which are probably attributed to the development of inter-grain Josephson coupling within the polycrystalline sample.~\cite{Yang1992.PhysicaC.201.325}
For SrRuO\sub{3}, a pronounced peak in $\varDelta V_y\sim \chi = \mathrm{d}M/\mathrm{d}H$ due to ferromagnetic ordering is observed at $162\pm 1$~K, in agreement with the known ferromagnetic transition temperature $T\sub{FM}\simeq 160$~K.~\cite{Longo1968.JApplPhys.39.1327}
In addition, a broad peak centered at around 100~K is also observed, which is attributable to glassy nature of the magnetic moments.~\cite{Sow2012.PhysRevB.85.224426}

\begin{figure}[tb]
\begin{center}
\includegraphics[bb=0.000000 0.000000 360.000000 252.000000,width=8.5cm]{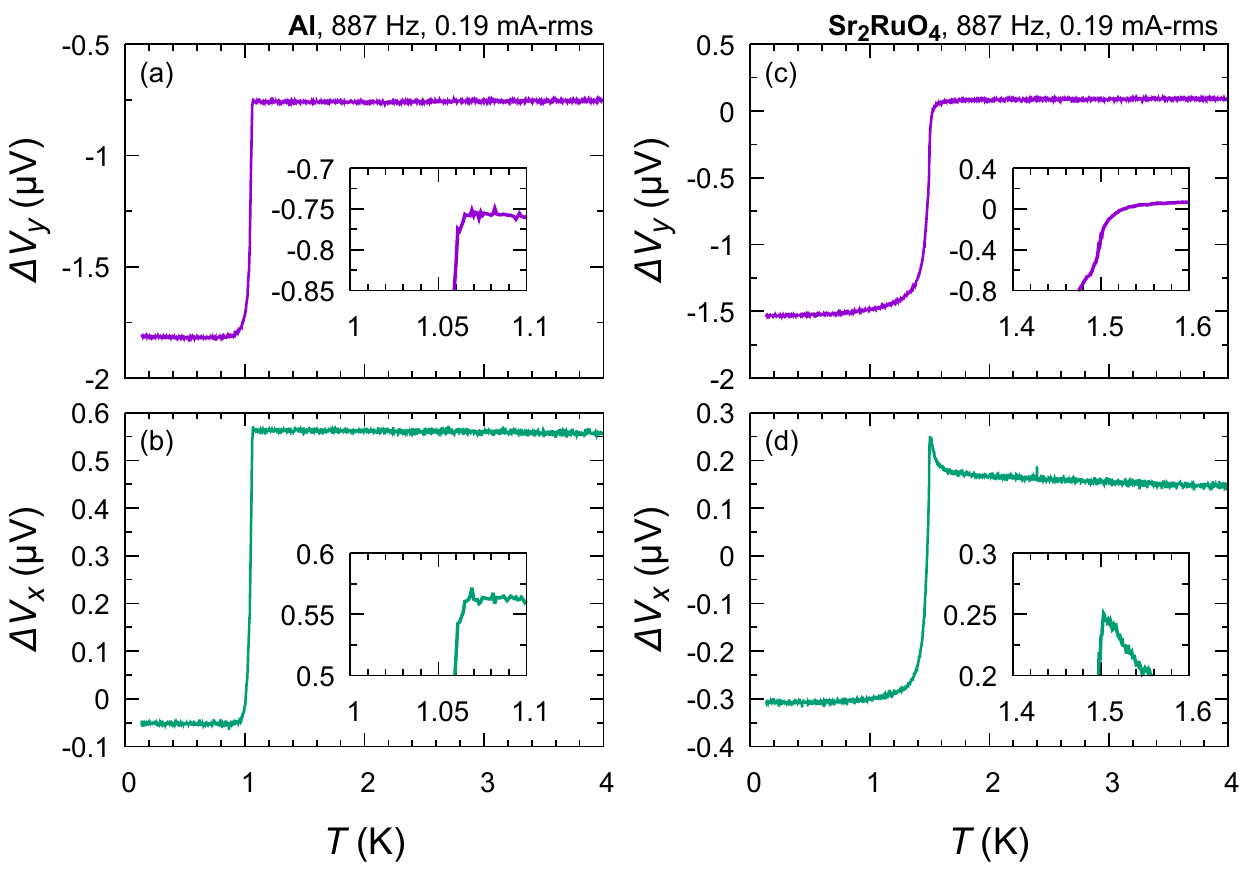}
\caption{Susceptometer responses of a polycrystalline sample of pure aluminum (12.06~mg) (panels (a) and (b)) and a single crystalline sample of \sro\ (21.79~mg) (panels (c) and (d)) measured with the AC current of 0.19~mA-rms ($\sim 6.3$~$\muup$T-rms) and 887~Hz.
The panels (a) and (c) present $\varDelta V_y$ after background subtraction for each sample, whereas (b) and (d) present $\varDelta V_x$.
The insets are enlarged view near $\Tc$ of the samples.
All data are collected in temperature up sweeps.
\label{fig:Al-SRO214}
}
\end{center}
\end{figure}

Figure~\ref{fig:Al-SRO214} presents susceptometer responses of polycrystalline pure aluminum (99.99\%, The Nilaco Corporation) and single crystalline \sro.
Since $\Tc$ of these materials are below 1.8~K, we used the ADR option (see Fig.~\ref{fig:mounting}(f)) to cool down the samples.
The \sro\ sample was grown with the floating-zone method.~\cite{Mao2000.MaterResBull.35.1813}
For aluminum, we applied a small negative field ($\mu_0H\sub{DC}=-0.0125$~T) to minimize the remnant field in the sample space.~\footnote{Our PPMS is equipped with a horizontal split magnet. Thus, the remnant field may be larger than that in an ordinary vertical magnet.}
Clear superconducting transitions were observed at 1.07~K for aluminum  and 1.50~K for \sro.
Although the latter exactly matches with the ideal $\Tc$ of this oxide in the clean limit,~\cite{Mackenzie1998.PhysRevLett.80.161} the former is slightly lower than the literature value~\cite{Phillips} ($\Tc=1.16$~K) due to a small uncancelled remnant magnetic field of the order of 0.001~T.
We emphasize here that it typically takes only 100~min to cool down from room temperature to 0.1~K with the ADR option (see Fig.~\ref{fig:T-time} in Appendix).
Thus, given the fast sample mounting process of our susceptometer, the combination of the ADR option and the present susceptometer provides an easy and fast method to check $\Tc$ of samples such as \sro, whose behavior has been revealed to be extremely sensitive to the sample quality, in particular near the upper critical field.~\cite{Yonezawa2013.PhysRevLett.110.077003}

\begin{figure}[tb]
\begin{center}
\includegraphics[bb=0.000000 0.000000 360.000000 252.000000,width=8.5cm]{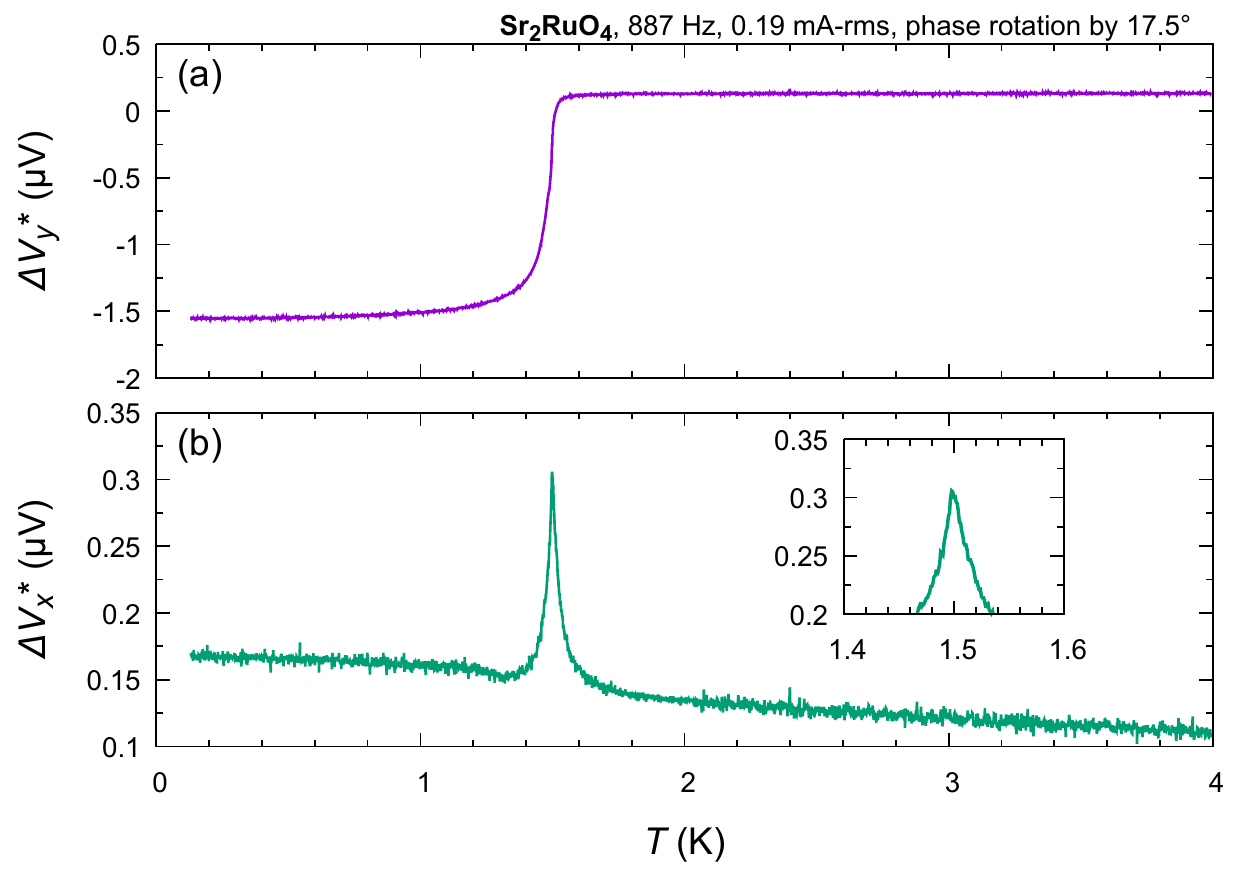}
\caption{Susceptometer response $\varDelta V_y^\ast$ and $\varDelta V_x^\ast$ of a \sro\ single crystal after background subtraction and phase rotation by $\theta = 17.5\deg$ (See Sec.~\ref{sec:phase}). The inset in (b) is an enlarged view of $\varDelta V_x^\ast(T)$ near $\Tc$.
\label{fig:SRO214_phase}
}
\end{center}
\end{figure}

The present data seem to have finite phase mixing between $\varDelta V_x$ and $\varDelta V_y$: 
although the background-subtracted $\varDelta V_y$ is nearly proportional to the real part of $\chiac$,
$\varDelta V_x$ contain information both the real and imaginary parts.
In the present case, the phase mixing is likely to depend on temperature and on sample, partly due to eddy current induced in the base plate and sample itself. 
The accurate determination of the phase correction angle $\theta$ (see Sec.~\ref{sec:phase}) is not so straightforward.
Nevertheless, phase correction is worth trying.
In Fig.~\ref{fig:SRO214_phase}, we plot phase-rotated values $\varDelta V_y^\ast$ and $\varDelta V_x^\ast$ with $\theta=17.5\deg$.
This $\theta$ value is determined so that the peak in $\varDelta V_x^\ast$ becomes most pronounced and the change in $\varDelta V_y^\ast$ becomes largest.
Although this value of $\theta$ may not be thoroughly accurate, $\varDelta V_x^\ast$ seems to be dominated by the imaginary part of $\chiac$.

\begin{figure}[tb]
\begin{center}
\includegraphics[bb=0.000000 0.000000 360.000000 252.000000,width=8.5cm]{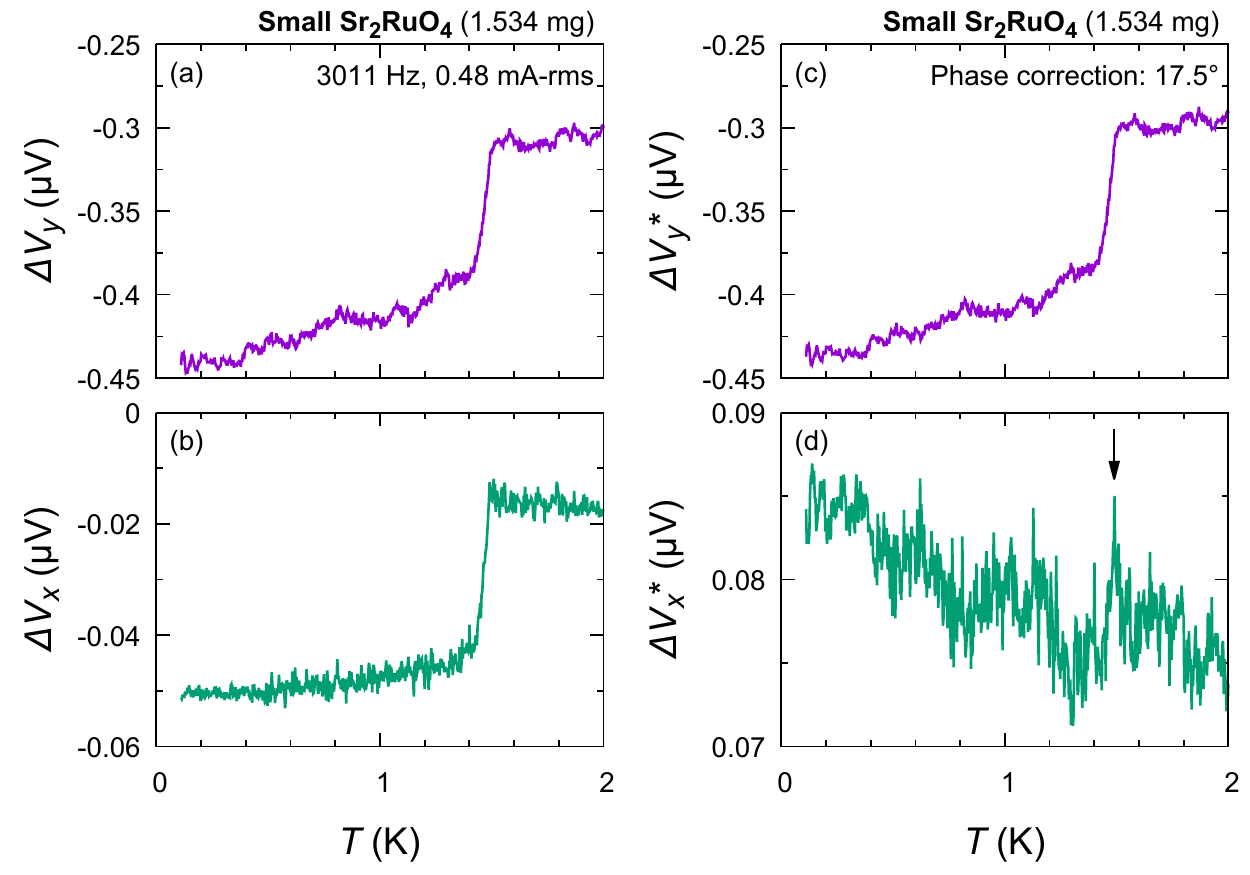}
\caption{
Susceptometer responses of a small single crystal of \sro\ (1.534~mg) measured with the AC current of 0.50~mA-rms ($\sim 17$~$\muup$T-rms) and 3011~Hz.
The panels (a) and (b) respectively present $\varDelta V_y$ and $\varDelta V_x$ after background subtraction.
In the panels (c) and (d), phase rotated values $\varDelta V_y^\ast$ and $\varDelta V_x^\ast$ with $\theta = 17.5\deg$ are shown. 
The arrow indicates the peak in $\varDelta V_x^\ast$ due to the SC transition at $\Tc = 1.49$~K.
\label{fig:SRO214_small-cryst}
}
\end{center}
\end{figure}

To check the sensitivity of the susceptometer for smaller samples, we measured SC response of tiny \sro\ single crystals.
In Fig.~\ref{fig:SRO214_small-cryst}, we present data of a single crystal with 1.534~mg ($\sim 1.0 \times 0.6 \times 0.4$~mm${}^3 = 0.24$~mm${}^3$) measured with AC current of 3011~Hz and 0.50~mA-rms ($\sim 17$~$\muup$T-rms).
Here, to compensate the signal reduction due to the small sample volume, we increased the amplitude and frequency of the AC field compared with previous mesurements.
As a result, we observed clear SC signal below $\Tc \sim 1.49$~K.
The phase rotated values with $\theta = 17.5\deg$ are also plotted.
A small peak in $\varDelta V_x^\ast$ at $\Tc$ can be seen as indicated with the arrow.
From these data, we estimate the minimal detectable SC sample volume as $\sim 1/5$ of the volume of this sample, namely $\sim 0.05$~mm${}^3$.
To further improve the sensitivity, one needs to modify the design so that the secondary coil has more turns and higher sample filling factor.
Such modifications would not be too difficult.
Use of a preamplifier to magnify the susceptometer signal is also worth trying.

\section{Summary}

To summarize, we developed a new design of an AC magnetic susceptometer that fits the PPMS and its ADR option.
With the susceptometer, in particular together with the ADR option, one can quickly investigate magnetic properties of samples down to 0.1~K, without much complicated preparation procedures. 
We emphasize that the design concept is applicable to susceptometers to be used in other apparatus than the PPMS.
This susceptometer helps researchers to accelerate process of improving sample quality as well as searching for new interesting materials.

\section*{Acknowledgements}
 
We acknowledge Quantum Design Japan, in particular T. Ohta, for their cooperation.
We also thank Y.~Kasahara, Y.~Yasui, M.~P.~Jimenez Segura for their technical assistance.
This work was supported by MEXT Grant-in-Aids for
Scientific Research on Innovative Areas on ``Topological Quantum Phenomena'' (KAKENHI 22103002) and ``Topological Materials Science''  (KAKENHI 15H05852), as well as by JSPS Grant-in-Aids KAKENHI 26247060 and 26287078.

\appendix

\section{Typical performance of the PPMS ADR option}

\begin{figure}[bt]
\begin{center}
\includegraphics[bb=0.000000 0.000000 360.000000 324.000000,width=8.5cm]{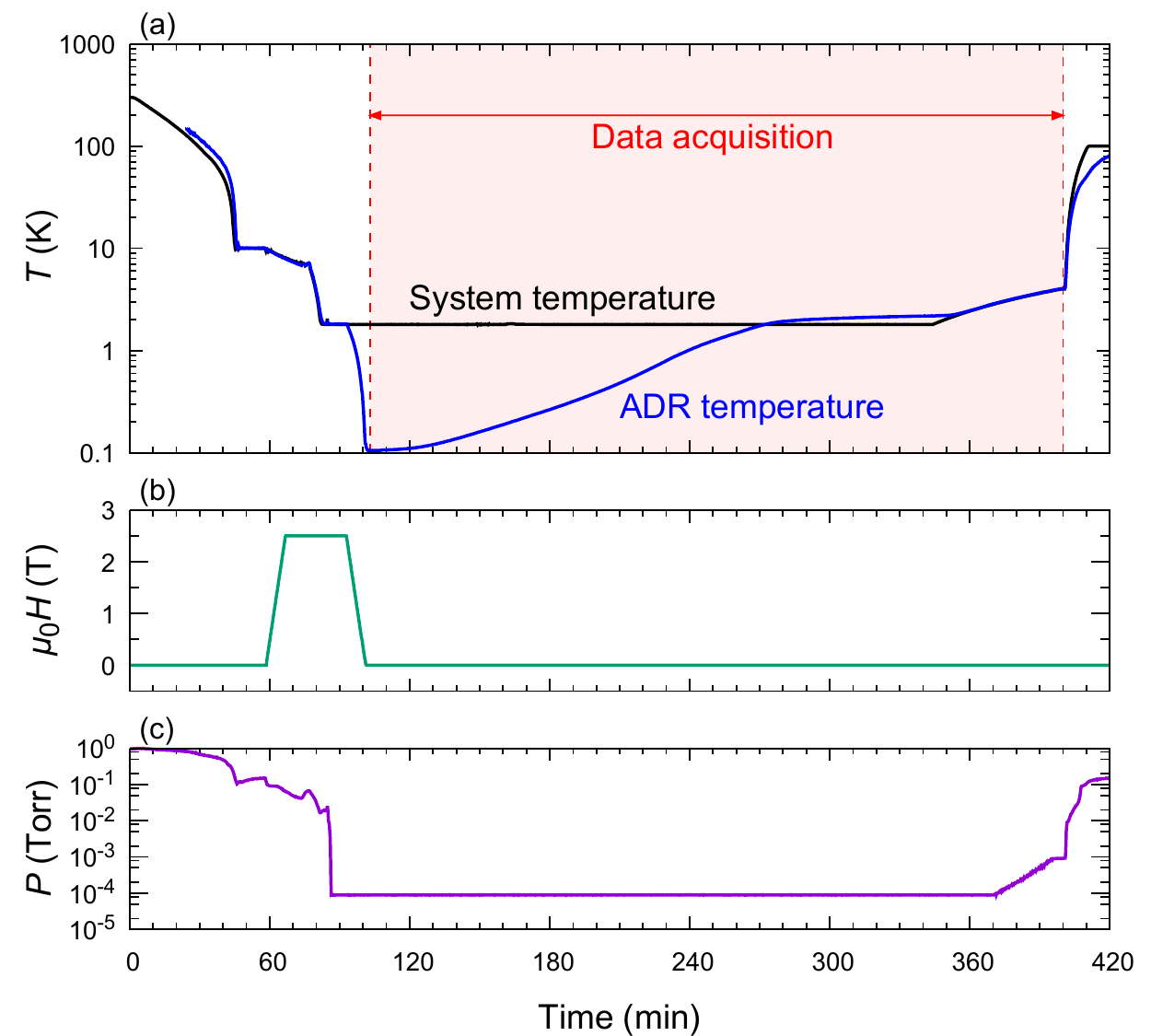}
\caption{Typical time dependence of various PPMS system parameters when using the ADR option.
The time evolution of the ADR temperature (blue curve) and PPMS system temperature (black curve) are plotted in the panel (a). 
It typically takes 100~min from 300~K to 0.1~K. 
Data are acquired in the time window indicated with the pink rectangle.
The changes of the external field and the chamber pressure are plotted in the panels (b) and (c), respectively.
\label{fig:T-time}
}
\end{center}
\end{figure}

In general, an ADR makes use of the entropy changes in a group of localized paramagnetic moments caused by external magnetic field.
After aligning the moments with a strong magnetic field, the magnetic field is switched off under adiabatic condition.
Heat is absorbed by the magnetic moment system when the magnetic moments becomes back to random in decreasing magnetic field.
In temperature ranges from a few Kelvin to sub Kelvin, paramagnetic salts are often used for this method.
For the ADR option of the PPMS, chromium alum is used.~\cite{ADR.PressRelease}

Figure~\ref{fig:T-time} represents typical time evolution of various parameters when we measured the AC susceptibility in the range 0.1~K${}<T<{}$ 4.0~K using the ADR option.
After the system temperature is stabilized at 10~K, the magnetic field is applied to 2.5~T and the system is cooled further down to 1.8~K. 
Then the high-vacuum pump is started to achieve adiabatic condition.
After pumping the chamber  for several minutes, the magnetic field is switched off to cool the sample down to 0.1~K.
It typically takes only 100~min from room temperature to 0.1~K.
Data is collected while the sample warms up naturally to $\sim 1.8$~K due to heat leak from outside.
This warming takes a few hours.
If one requires data above 1.8~K, the system temperature is slowly increased to the target temperature after the ADR temperature reaches 1.8~K.

\section{Raw signal and background}
\label{sec:background}

\begin{figure}[tb]
\begin{center}
\includegraphics[bb=0.000000 0.000000 360.000000 252.000000,width=8.5cm]{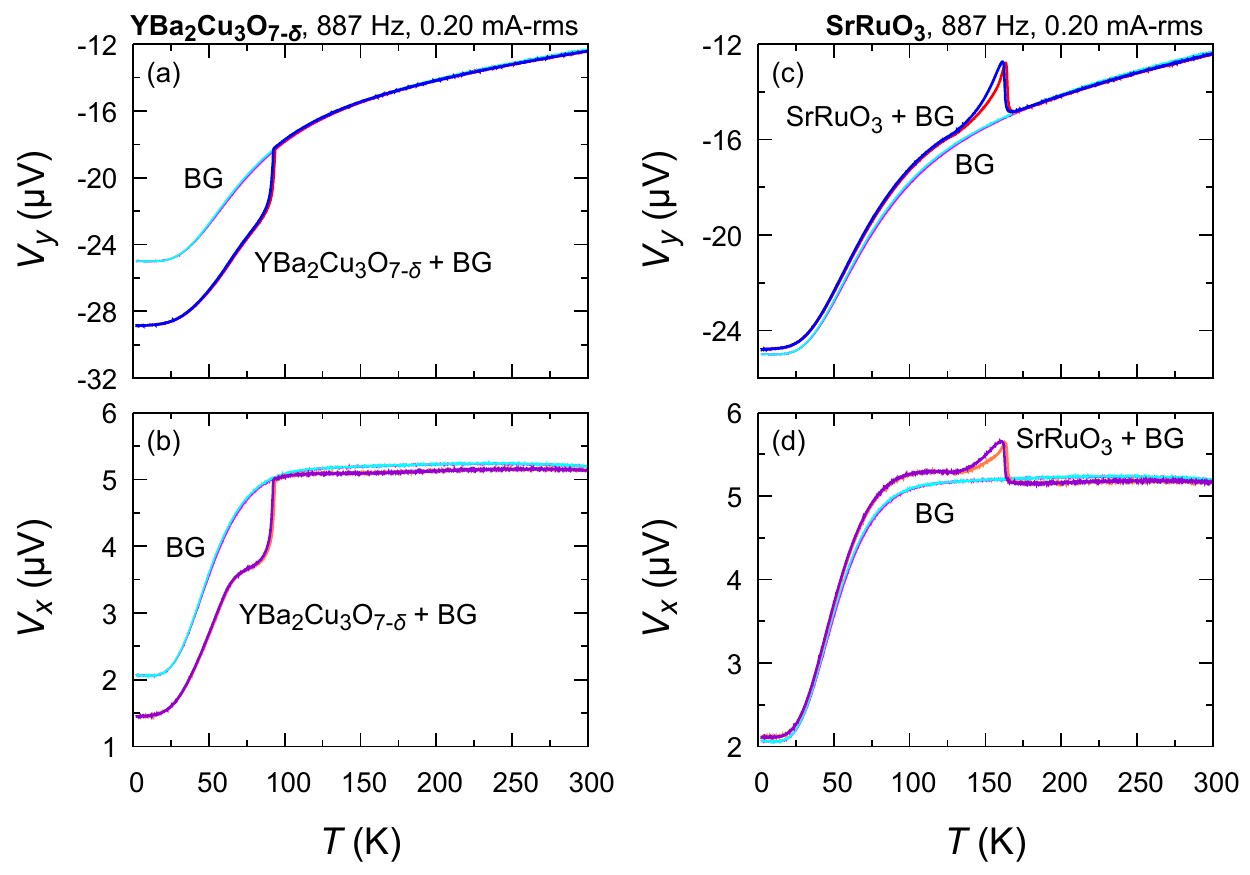}
\caption{Output voltages of the susceptometer of YBCO and SrRuO\sub{3} polycrystals, compared with the background signals.
\label{fig:YBCO_SRO113_raw}
}
\end{center}
\end{figure}

In Fig.~\ref{fig:YBCO_SRO113_raw}, we compare the output voltage of the susceptometer with the background signals measured with the ordinary sample chamber of the PPMS.
Although the temperature dependence of the background contribution was rather strong at temperatures below 100~K, clear diamagnetic signals originating from superconductivity of YBCO or the ferromagnetic transition of SrRuO\sub{3} were observed. 
The background is quite reproducible and can be easily subtracted from the sample data by fitting the background with a polynomial function.

\begin{figure}[bt]
\begin{center}
\includegraphics[bb=0.000000 0.000000 360.000000 252.000000,width=8.5cm]{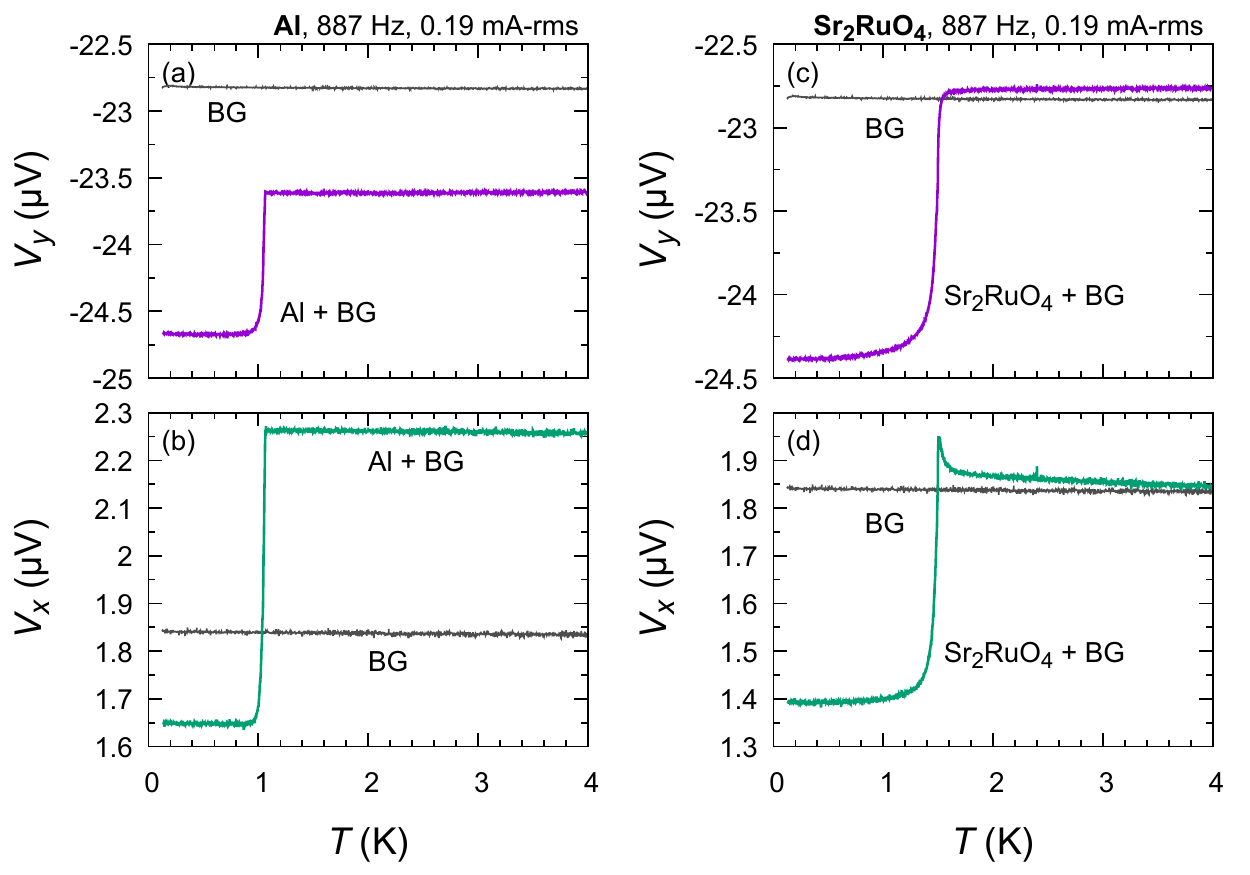}
\caption{Output voltage of the susceptometer of Al and \sro, compared with the background signals.
\label{fig:Al_SRO214_raw}
}
\end{center}
\end{figure}

For low-temperature measurements using the ADR option, the temperature dependence of the background is very small as shown in Fig.~\ref{fig:Al_SRO214_raw},
except for a step-like anomaly at 9.2~K originating from superconducting transition of Nb-Ti wire used in the ADR (not shown).
Notice that the offset between the background and the sample plus background signals above $\Tc$ is quite large for aluminum, despite it is a simple metal.
In contrast, for \sro, the two data nearly coincide with each other above $\Tc$. 
The observed large offset in the normal state of aluminum must originate from the eddy current response, which can be significant in highly conductive materials.
Indeed, the skin depth $\delta\sub{s}$ of aluminum is estimated to be $\sim 17~\muup$m using the relation $\delta\sub{s} = \sqrt{\rho/\pi f \mu}$, where $\rho\sim 1\times 10^{-12}$~$\Omega$m is the resistivity of aluminum below 4~K,~\cite{Desai1984.JPhysChemRefData.13.1131} $f = 887$~Hz is the measurement frequency, and $\mu\sim 4\pi \times 10^{-7}$~Wb/Am is the permeability of aluminum.
Thus, considering the sample size of a few mm, the AC magnetic field is already expelled from a large fraction of the sample even above $\Tc$.

\section{Phase correction}
\label{sec:phase}

In the mutual inductance method, it is ideally expected that $\varDelta V_y \propto \chi'$ and $\varDelta V_x \propto \chi''$, if one measures $V_x$ and $V_y$ with respect to the phase of the AC current. The phase of the current can be determined from the voltage across the standard register (see Fig.~\ref{fig:electronics}).
In reality, however, there can be a mixing between $\varDelta V_y$ and $\varDelta V_x$:
The observed $\varDelta V_y$ and $\varDelta V_x$ can be given by 
\begin{align}
\begin{pmatrix}
\varDelta V_y \\
\varDelta V_x
\end{pmatrix}
= 
A
\begin{pmatrix}
\cos\theta & -\sin\theta \\
\sin\theta & \cos\theta 
\end{pmatrix}
\begin{pmatrix}
\chi' \\
\chi''
\end{pmatrix},
\end{align}
where $A$ is a coefficient and $\theta$ is the phase mixing angle.

Once the value of $\theta$ is determined, which is actually not so straightforward in some cases, we can obtain the phase-rotated values $\varDelta V_y^\ast$ and $\varDelta V_x^\ast$ defined by
\begin{align}
\begin{pmatrix}
\varDelta V_y^\ast \\
\varDelta V_x^\ast
\end{pmatrix}
= 
\begin{pmatrix}
\cos\theta & \sin\theta \\
-\sin\theta & \cos\theta 
\end{pmatrix}
\begin{pmatrix}
\varDelta V_y \\
\varDelta V_x
\end{pmatrix}.
\end{align}
In some cases, $\theta$ is determined so that the peak in $\varDelta V_x^\ast$ becomes most pronounced and the change in $\varDelta V_y^\ast$ becomes largest.

\section{Field distribution in the primary coil}
\label{sec:field-distribution}

The field/current ratio of the primary coil is calculated using the formula in the SI unit
\begin{align}
H(z)=\frac{NI}{2l}\left[ \frac{z+l/2}{\sqrt{a^2 + (z + l/2)^2}} - \frac{z-l/2}{\sqrt{a^2 + (z - l/2)^2}} \right],
\end{align}
where $z$ is the position along the center axis of the coil with the origin at the middle of the coil, $N = 400$ is the total number of turns of the coil, $a = 5$~mm is the radius of the coil, and $l=7$~mm is the length of the coil.
We evaluate the ratio to be $\mu_0 H/I \sim 0.033$~T/A at the sample position ($z\sim -2.7$ mm).
We comment here that the ratio rather strongly depends on $z$ because of the short aspect ratio of the present coil.
Thus, the field actually felt by the sample is not homogeneous if the sample is rather large.


\bibliography{%
../AC-chi.bib,%
../../../paper/string.bib,%
../../../paper/TMTSF,%
../../../paper/organic_SC,%
../../../paper/textbook,%
../../../paper/FFLO,%
../../../paper/CeCoIn5,%
../../../paper/Type-I_SC,%
../../../paper/heavy-Fermion,%
../../../paper/superconductors,%
../../../paper/SC,%
../../../paper/Sr2RuO4,%
../../../paper/Sr3Ru2O7,%
../../../paper/high-Tc,%
../../../paper/oxides,%
../../../paper/measurement_technique,%
../../../paper/Fe-Pn%
}






\end{document}